\documentstyle[12pt]{article}

\begin{document}

\setcounter{page}{1}

\pagestyle{plain} \vspace{1cm}
\begin{center}
\Large{\bf Noncommutative Geometry Inspired Entropic Inflation}\\
\small \vspace{1cm} {\bf Kourosh Nozari$^{a,b,*}$}\quad and\quad {\bf Siamak Akhshabi$^{a,\dagger}$}\\
\vspace{0.5cm} {\it $^{a}$Department of Physics,
Faculty of Basic Sciences,\\
University of Mazandaran,\\
P. O. Box 47416-95447, Babolsar, IRAN\\
\vspace{0.25cm} $^{b}$ Research Institute for Astronomy and
Astrophysics of Maragha,\\
P. O. Box 55134-441, Maragha, IRAN\\
\vspace{0.25cm}
$^{*}$knozari@umz.ac.ir\\
$^{\dagger}$s.akhshabi@umz.ac.ir}

\end{center}
\vspace{1.5cm}
\begin{abstract}
Recently Verlinde proposed that gravity can be described as an
emergent phenomena arising from changes in the information
associated with the positions of material bodies. By using
noncommutative geometry as a way to describe the microscopic,
microstructure of quantum spacetime, we derive modified Friedmann
equation in this setup and study the entropic force modifications
to the inflationary dynamics of early universe.
\end{abstract}
\vspace{2cm}
\newpage
\section{Introduction}
Recently, Verlinde [1] proposed an idea similar to Jacobson's
thermodynamic derivation of the Einstein equations [2], where it is
argued that Newton's law of gravitation can be understood as an
entropic force caused by information changes when a material body
moves away from the holographic screen. There are some interesting
consequences arising from this argument. One is that with the
assumption of the entropic force together with the Unruh temperature
[3], Verlinde is able to derive the second law of Newton. The other
is that the assumption of the entropic force together with the
holographic principle and the equipartition law of energy leads to
Newton's law of gravitation. Consequences of Verlinde's proposal in
cosmology  have been widely studied [4,5,6]. The inflationary
dynamics of a universe governed by entropic gravity also has been
studied [7].

To derive the Newton's law from thermodynamics, Verlinde considered
a system consisting of two masses, one test mass $m$ and a source
mass $M$. One can consider a surface $\Omega$ centered around $M$
and lying between the two masses. One also assumes $m$ to be at a
distance from $\Omega$ smaller with respect to its reduced Compton
wavelength $\lambda_{m} = \hbar/(mc)$. Verlinde then proposed that
when the test particle moves in the vicinity of the holographic
screen $\Omega$, the change in entropy is proportional to the
displacement $\Delta x$ i.e.
\begin{equation}
\Delta S_{\Omega}=2\pi k_{B}\frac{\Delta x}{\lambda_{m}}\,.
\end{equation}
Number of bits of information stored in the surface of $\Omega$ is
given by
\begin{equation}
N=\frac{A_{\Omega}}{l^{2}_{p}}
\end{equation}
where $l_{p}$ is the Planck length. The surface of $\Omega$ is in
thermal equilibrium at the temperature $T$, all bites are equally
likely and the energy of $\Omega$ is equipartitioned among them i.e.
\begin{equation}
U_{\Omega}=\frac{1}{2}Nk_{b}T
\end{equation}
where $U_{\Omega}$\, is equal to the rest mass of the source
$Mc^{2}$. From thermodynamical equation of this system, a force $F$
will arise
\begin{equation}
F\Delta x=T\Delta S_{\Omega}\,.
\end{equation}
From the above argument, one can reach the gravitational law of
Newton i.e $F=GMm/r^{2}$, (see [1] for more details).

\section{Noncommutative geometry inspired entropic correction to Newton's law}
To link this entropic interpretation of gravity with noncommutative
geometry, one should consider  entropy as a tool to  connect the
standard description of gravity with the underlying microstructure
of a quantum spacetime. Here we use noncommutative geometry to
describe the microscopic structure of spacetime manifold. Inspired
by some aspects of string theory and loop quantum gravity, the basic
idea of noncommutative geometry is that due to existence of a
fundamental minimal length (Planck or string length), the coordinate
operators also fail to commute. In this case, the  \emph{fuzziness}
of spacetime can be expressed using the following relation for
non-commutativity of coordinate operators [8,9]
\begin{equation}
[\hat{x}^i,\hat{x}^j]=i\theta^{ij}
\end{equation}
where $\theta^{ij}$ is a real, antisymmetric matrix, with the
dimension of length squared which determines the fundamental cell
discretization of spacetime manifold. As a consequence of above
relation, the notion of point in the spacetime manifold becomes
obscure as there is a fundamental uncertainty in measuring the
coordinates
\begin{equation}
\Delta x^{i}\Delta x^{j}\geq \frac{1}{2}|\theta^{ij}|.
\end{equation}
The presence of noncommutativity enforces some modifications on
Verlinde assumptions [10]. First due to uncertainty on the surface
$\Omega$, there exists a fundamental unit for entropy change,
$\Delta S_{\theta}$ which is realized at the displacement $\Delta
x_{min}$. Therefore the change of entropy is
\begin{equation}
\Delta S_{\Omega}=\Delta S_{\theta}(\frac{\Delta x}{\Delta
x_{min}})\,.
\end{equation}
This equation states that when the test mass $m$ is a distance
$\Delta x_{min} $ away from the surface, the entropy of the surface
changes by one fundamental unit $\Delta S_{\theta}$. In other words,
due to existence of a minimal length, $\Delta x_{min}$, there is a
minimum change of entropy as $\Delta S_{\theta}$. A general
displacement $\Delta x$ then results in a change in entropy as
equation (7), where $(\frac{\Delta x}{\Delta x_{min}})$ is an
integer number.

Secondly, on the surface $\Omega$ the unit surface is $\theta$. So
the number of information bits is
\begin{equation}
N=\frac{A_{\Omega}}{\theta}\,.
\end{equation}
Using these assumptions, the entropic modification to Newton's law
has been derived in [10] as
\begin{equation}
F=\frac{Mm}{r^{2}}\Big(\frac{4c^{3}\theta^{2}}{\hbar\alpha^{2}}\Big)\bigg[\frac{c^{3}}{4\hbar
G}+\frac{\partial s}{\partial A}\bigg]
\end{equation}
where $\alpha$ is defined as
\begin{equation}
\alpha=\frac{8\pi \Delta x_{min}}{\lambda_{m}}\,.
\end{equation}
For the entropic force in (9) to coincide with the Newton's law, we
should set the noncommutative parameter $\theta$ as $\theta=\alpha
l^{2}_{p}$. Doing this, the modified Newton's law reads
\begin{equation}
F=\frac{GMm}{r^{2}}\bigg[1+4l^{2}_{p}\frac{\partial s}{\partial
A}\bigg]\,,
\end{equation}
which coincides with the Newton's law to the first order. One should
note that the second term in the bracket is a quantum correction
which arises when one considers quantum gravity corrections to the
entropy-area relation [11]. If we neglect the quantum effects, the
second term in (11) vanishes.

With the same assumptions, the temperature of the noncommutative
holographic screen will be [10]
\begin{equation}
T=\frac{M}{r^2}\frac{\theta c^2}{2\pi k_{B}}\,.
\end{equation}
From this equation, it is obvious that a noncommutative manifold is
equivalent to a thermodynamical system whose temperature is given by
the above relation. \\

To obtain the  explicit modification of Newton's law caused by
entropic corrections, we should have the specific entropy-area
relation in eq. (11). To achieve this goal, we use the coherent
state picture of noncommutativity first proposed in Ref. [12] as our
underlying microstructure of a quantum spacetime. This approach
eliminates point-like structures in the favor of smeared objects.
Using this approach for obtaining the entropy/area relation with the
help of black hole thermodynamics, the modified Newton's law has
been derived in Ref. [10] as
\begin{equation}
F=\frac{GMm}{r^{2}}\bigg[1+\frac{\Gamma(3/2;(r^{2}/4\pi\theta))}{\gamma(3/2;(r^{2}/4\pi\theta))}\bigg]
\end{equation}
where $\Gamma(3/2;(r^{2}/4\pi\theta))$ and
$\gamma(3/2;(r^{2}/4\pi\theta))$ are incomplete upper and lower
gamma functions respectively. In the low energy or large distance
limit( $r\gg\theta$ ), this reduces to
\begin{equation}
F\simeq \frac{GMm}{r^{2}}\bigg[1+\frac{re^{-r^{2}/(4\alpha
l^{2}_{p})}}{\sqrt{\pi\alpha}l_{p}}+\frac{r^{2}e^{-r^{2}/(2\alpha
l^{2}_{p})}}{2\alpha\sqrt{\pi}l^{2}_{p}}\bigg]\,.
\end{equation}
In the next section we will use this modified force to obtain the
cosmological dynamics of the universe using the approach of
Newtonian cosmology.

\section{Cosmological dynamics}
To obtain cosmological dynamics in this scenario, we assume that the
background spacetime is spatially homogeneous and isotropic and is
given by the Friedmann-Robertson-Walker (FRW) metric
\begin{equation}
ds^{2}=h_{ij}dx^{i}dx^{j}+R^{2}\bigg(d\vartheta^{2}+\sin^{2}(\vartheta)d\varphi^{2}\bigg)\,,
\end{equation}
where $i,\,j=0,\,1$. We consider a region of spacetime with volume
$V$ with a compact boundary surface $S$. The region is a sphere with
radius $R=a(t)r$ where $r$ will remain constant during expansion.
Because of the spherical symmetry, this boundary is an equipotential
surface such that we may treat it as a holographic screen. Now we
assume that in the Newtonian cosmology the background spacetime is
Minkowskian. For this region the apparent (Hubble) horizon will be
at $R=1/H$. So, we assume that in this setup the holographic screen
is at the apparent horizon $R=1/H$ for a flat spacetime. A similar
approach could be found in references [4,5,13]. Based on this
argument, the dynamical apparent horizon, a marginally trapped
surface with vanishing expansion, is given in a FRW background by
\begin{equation}
R=ar=\frac{1}{\sqrt{H^{2}+k/a^{2}}}\,.
\end{equation}
We assume also that matter source in the FRW universe is a perfect
fluid with stress-energy tensor
\begin{equation}
T_{\mu\nu}=(\rho+p)u_{\mu}u_{\nu}+pg_{\mu\nu}\,.
\end{equation}
The system also obeys the usual continuity equation as
\begin{equation}
\dot{\rho}+3H(\rho+p)=0\,.
\end{equation}
With these assumptions and using the modified force (14), now we
find the acceleration equation following the procedures of Newtonian
cosmology.  We consider a region of spacetime with volume $V$ with a
compact surface $S$. The region is a sphere with radius $R=a(t)r$
where $r$ remains constant during expansion. Applying Newton's
second law and using the force as given in (14) for a test mass near
the surface $S$, we get
\begin{equation}
F=ma=m\ddot{R}=mr\ddot{a}=-\frac{GMm}{r^{2}}\bigg[1+\frac{re^{-r^{2}/(4\alpha
l^{2}_{p})}}{\sqrt{\pi\alpha}l_{p}}+\frac{r^{2}e^{-r^{2}/(2\alpha
l^{2}_{p})}}{2\alpha\sqrt{\pi}l^{2}_{p}}\bigg]\,.
\end{equation}
Given that $\rho=M/V$ and $V=\frac{4}{3}\pi a^{3}r^{3}$, equation
(19) can be rewritten as
\begin{equation}
\frac{\ddot{a}}{a}=-\frac{4\pi
G}{3}\rho\bigg[1+\frac{Re^{-R^{2}/(4\alpha
l^{2}_{p})}}{\sqrt{\pi\alpha}l_{p}}+\frac{R^{2}e^{-R^{2}/(2\alpha
l^{2}_{p})}}{2\alpha\sqrt{\pi}l^{2}_{p}}\bigg]\,.
\end{equation}
To obtain the correct Friedmann equation of the model, we note that
it is the active gravitational mass ${\cal{M}}$ in the spatial
region $V$ rather than the total mass $M$ that produces the
acceleration [4]. The active gravitational mass or Tolman-Komar mass
[14], is defined as
\begin{equation}
{\cal{M}}=2\int dV
(T_{ab}-\frac{1}{2}Tg_{ab})u^{a}u^{b}=(\rho+3p)\frac{4}{3}\pi
a^{3}r^{3}
\end{equation}
Replacing $M$ with ${cal{M}}$ in (19) we get
\begin{equation}
\frac{\ddot{a}}{a}=-\frac{4\pi
G}{3}(\rho+3p)\bigg[1+\frac{Re^{-R^{2}/(4\alpha
l^{2}_{p})}}{\sqrt{\pi\alpha}l_{p}}+\frac{R^{2}e^{-R^{2}/(2\alpha
l^{2}_{p})}}{2\alpha\sqrt{\pi}l^{2}_{p}}\bigg]\,.
\end{equation}
Using the above acceleration equation and the continuity equation
(18), we find the Friedmann equation as follows
\begin{equation}
H^{2}+\frac{k}{a^{2}}=\frac{8\pi
G}{3}\rho\bigg[1+\frac{Re^{-R^{2}/(4\alpha
l^{2}_{p})}}{\sqrt{\pi\alpha}l_{p}}\int\frac{d(\rho
a^{2})}{a^{2}}+\frac{R^{2}e^{-R^{2}/(2\alpha
l^{2}_{p})}}{2\alpha\sqrt{\pi}l^{2}_{p}}\int\frac{d(\rho
a^{2})}{a^{4}}\bigg]\,.
\end{equation}

If we assume a constant equation of state parameter, then the energy
density will be given by $\rho=\rho_{p}a^{-3(1+\omega)}$ where
$\rho_{p}$ is an integration constant corresponding to present value
of the energy density. In this case the integrals in Friedmann
equation (23) can be solved analytically to find
\begin{equation}
{H}^{2}+{\frac {k}{{a}^{2}}}=\frac{8\pi \,G}{3}\rho\Bigg[ 1+\frac{(
1+3\,\omega )}{3( 1+\omega )\sqrt {\pi \,\theta}}\rm R{{\rm
e}^{-{\frac {{R}^{2}}{4\theta}}}}+\frac{( 1+3\,\omega )}{( 5+3\omega
)\sqrt {\pi }\theta}\rm R^2{{\rm e}^{-{\frac {{R}^{2}}{2\theta}}}}
\Bigg]\,.
\end{equation}

The existence of exponential terms in the acceleration equation has
interesting consequences at very early times in the inflation era.
First of all, it should be noted that the terms in the brackets of
equation (24) ( and similarly in equation (22)) are positive for
ordinary matter fields, so we still need $\rho+3p<0$ to have
inflation. Fortunately, noncommutativity provides the negative
pressure needed for a successful inflation. To examine this, we use
a newly proposed model for noncommutative inflation introduced in
[15]. The basic idea is that the initial singularity is smeared due
to spacetime noncommutativity.  One could split the energy density
on any hypersurface as [15]
\begin{equation}
\rho=\rho_{0}e^{-|T|^{2}/4\theta}e^{-|\vec{{X}}|^{2}/4\theta}
\end{equation}
where $R^{2}=T^{2}+|\vec{X}|^{2}$ and $T=it$ is the Euclidean time.
From one hypersurface to another, the $\vec{X}$-dependent part of
$\rho$ does not change, so it can be included into $\rho_{0}$.
Therefore, one could write the energy density of the smeared initial
singularity as [10]
\begin{equation}
\rho(t)=\rho_{0}\, e^{-t^{2}/4\theta}\,.
\end{equation}
Using equation (16) for apparent horizon $R$, the Friedmann equation
(22) could be solved analytically to obtain the scale factor $a(t)$
as

$$a(t)\simeq\exp\bigg[\frac{4}{3}\pi \,\sqrt
{3G{\rho_{0}}\,\theta}\,\, {{\rm erf}({\frac {t}{2\sqrt
{2\theta}}})}\bigg]$$
\begin{equation}
 +{\frac {{{\rm e}^{{\frac {t}{2\sqrt {2\theta}}}}}}{256\sqrt{\pi}\theta^{2}}}\,
\bigg[\frac{\sqrt {2\pi}}{8}{t}^{5}{{\rm e}^{{\frac {t}{2\sqrt
{2\theta}}}}}{\theta}^{-5/2}+{\frac {25}{ 64}}\,{\rm Ei}(1,\,{\frac
{t}{2\sqrt {2\theta}}}
 ) {t}^{4}{{\rm e}^{{\frac {t}{2\sqrt {2\theta}}}}}{
\theta}^{-2}+{\frac {7}{8}}\,{\frac {{t}^{2}}{\theta}}+{\frac
{7\sqrt {2}}{32}} \,{\frac {{t}^{3}}{{\theta}^{3/2}}}+{\frac {t\sqrt
{2} }{2\sqrt {\theta}}}+6\bigg]\,,
\end{equation}
where  ${\rm \,Ei}(a,z)$ is the exponential integral defined as
${\rm \,Ei}(a,z)=z^{a-1}\Gamma(1-a,z)$.

Using the continuity equation (18), pressure is given by

$$p={\rho_{0}}\,{{\rm e}^{-{\frac {{t}^{2}}{4\theta} }}}\Bigg\{ 16384\,\sqrt {\pi
}t{\theta}^{5}\exp\Bigg[{\frac{4\sqrt {3}}{3}\pi \,\sqrt {G{
\rho_{0}}\,\theta}\,\, {{\rm erf}({\frac {t}{2\sqrt
{2\theta}}})}}\Bigg] +{{\rm e}^{{\frac {t }{\sqrt
{2\theta}}}}}\Bigg[8\,\sqrt {\pi }{t}^{6}\sqrt
{2\theta}+25\,{t}^{5}\theta\,{\rm Ei}( 1,{\frac {t}{2\sqrt
{2\theta}}})\Bigg]$$

$$+{ {\rm e}^{{\frac {t}{2\sqrt
{2\theta}}}}}\Bigg[164\,{t}^{3}{\theta}^{2}+14\,{t}^{4}{
\theta}^{3/2}\sqrt {2}-384\,{ \theta}^{3}t-304\,\sqrt
{2}{t}^{2}{\theta}^{5/2}\Bigg]$$

$$-65536\,\pi \,\sqrt {3}\sqrt {G{\rho_{0}}\,\theta}{\theta}^{11/2}\exp\Big[\frac{1}{24}(
-3{t}^{2}+32\pi \,\sqrt {6} \sqrt {G{\rho_{0}}\,\theta}\,\, {{\rm
erf}(\frac {t}{2\sqrt {2\theta}})} \Big]-{{\rm e}^{{\frac
{t}{\sqrt {2\theta}}}}}\Bigg[240\sqrt
{2\pi}\,{\theta}^{3/2}{t}^{4}$$

$$-48\theta\,{t}^{5}\sqrt {\pi }-600{\theta}^{2}{\rm
Ei}( 1,{\frac {t}{2\sqrt {2\theta}}}) {t}^{3}-75\,\sqrt {2}
{t}^{4}{\theta}^{3/2}{\rm Ei}( 1,{ \frac {t}{2\sqrt
{2\theta}}})\Bigg]$$

$$-768\,{ {\rm e}^{{\frac {t}{2\sqrt
{2\theta}}}}}{\theta}^{7/2} \sqrt {2}\Bigg\}
\times{\theta}^{-1}\times\Bigg\{ 65536\,\pi \,\sqrt {3G{\rho_{0}}\,
\theta}\exp\Big[{\frac{1}{24}( -3\,{t}^{2}+32\,\pi \,\sqrt {3}\sqrt
{G {\rho_{0}}\,\theta}\,\, {{\rm erf}({\frac {t}{2\sqrt
{2\theta}}})} ) }\Big]\sqrt {2}{\theta}^{9/2}$$

$$+{{\rm e}^{{\frac {t}{\sqrt {2\theta}}}}}\Bigg[240\,\sqrt
{2\pi\theta}{t}^{4}+48\,{t}^{5}\sqrt {\pi }+600\,\theta\,{\rm Ei}(
1,{\frac {t}{2 \sqrt {2\theta}}}) {t}^{3}+75\,\sqrt {\theta}{\rm
Ei}( 1,{\frac {t}{2\sqrt {2\theta}}}) {t}^{4}\sqrt {2}\Bigg]$$
\begin{equation}
-{{\rm e}^{{\frac {t }{2\sqrt
{2\theta}}}}}\Bigg[108\,{t}^{3}\theta\,+768\,{\theta}^{2}t+336\,\sqrt
{2}{t}^{2}{ \theta}^{3/2}+768\,\sqrt
{2}{\theta}^{5/2}\Bigg]\Bigg\} ^{-1}\,.
\end{equation}
And the equation of state parameter will be

$$\omega=- \Bigg\{ -16384\,\sqrt {\pi }t{\theta}^{5}\exp\Big[{\frac{4}{3}\pi \,\sqrt {3G{
\rho_{0}}\,\theta}\,\, {{\rm erf}({\frac {t}{2\sqrt
{2\theta}}})}}\Big] -8\,{t}^{6}\sqrt {\theta}\sqrt {2}{{\rm
e}^{1/2\,{\frac {t \sqrt {2}}{\sqrt {\theta}}}}}\sqrt {\pi
}-25\,{t}^{5}\theta\,{\rm Ei} ( 1,{\frac {t}{2\sqrt {2\theta}}} )
{{\rm e}^{ {\frac {t}{\sqrt {2\theta}}}}}$$

$$-{ {\rm e}^{{\frac
{t}{2\sqrt {2\theta}}}}}\Bigg[164\,{t}^{3}{\theta}^{2}-14\, \sqrt
{2}{t}^{4}{ \theta}^{3/2}+384\,{ \theta}^{3}t+304\,\sqrt
{2}{t}^{2}{\theta}^{5/2}\Bigg]$$

$$+65536\,\sqrt {2}\pi {\theta}^{11/2}\,\sqrt
{3G{\rho_{0}}\,\theta}\exp\Big[{\frac{1}{24}( -3\,{t}^{2}+32\,\pi
\, \sqrt {3G{\rho_{0}}\,\theta}\,\, {{\rm erf}({\frac {t}{2\sqrt
{2\theta}}})})}\Big]$$

$$+{{\rm e}^{ {\frac {t}{\sqrt
{2\theta}}}}}\Bigg[240\,{\theta}^{3/2}{t}^{4}\sqrt {2\pi
}+48\,\theta\,{t}^{5}\sqrt {\pi }+600\,{\theta}^{2}{\rm Ei}(
1,{\frac {t\sqrt { 2}}{4\sqrt {\theta}}}) {t}^{3}+75\,\sqrt
{2}{\theta}^{3/2}{\rm Ei}( 1,{ \frac {t}{2\sqrt {2\theta}}})
{t}^{4}\Bigg]$$
$$+768\,\sqrt
{2}{ {\rm e}^{{\frac {t}{2\sqrt {2\theta}}}}}{\theta}^{7/2}
\Bigg\}\times {\theta}^{-1}\times\Bigg\{ 65536\,\pi \,\sqrt {3G{
\rho_{0}}\,\theta}\exp\Big[{\frac{1}{24}( -3\,{t}^{2}+32\,\pi \,
\sqrt {3G{\rho_{0}}\,\theta}\,\, {{\rm erf}({\frac {t}{2\sqrt
{2\theta}}})})}\Big]\sqrt {2}{\theta}^{9/2}$$

$$+{{\rm e}^{{\frac {t}{\sqrt
{2\theta}}}}}\Bigg[240\,\sqrt {2\pi\theta}{t}^{4}+48\,{t}^{5}\sqrt
{\pi }+600\,\theta\,{\rm Ei}( 1,{\frac {t}{2\sqrt {2\theta}}})
{t}^{3}+75\,\sqrt {2}\sqrt {\theta}{\rm Ei}( 1,{\frac {t }{2\sqrt
{2\theta}}}) {t}^{4}\Bigg]$$
\begin{equation}
-{{\rm e}^{{\frac {t \sqrt {2}}{4\sqrt
{\theta}}}}}\Bigg[108\,{t}^{3}\theta\,+768\,{\theta}^{2}t+336\,\sqrt
{2}{t}^{2}{ \theta}^{3/2} +768\,\sqrt {2}{
\theta}^{5/2}\Bigg]\Bigg\} ^{-1}\,\,.
\end{equation}
Although these equations have very complicated form, the physical
implications of them are very simple. In figure 1 we show variation
of pressure and the equation of state parameter with the cosmic time
in the early stages of the universe evolution. The pressure is
negative which provides the condition for a successful inflation.
Also the plot of equation of state parameter versus the cosmic time
shows an accelerating behavior ( $\omega<-\frac{1}{3}$) in early
stage of the universe evolution.
\begin{figure}[htp]
\begin{center}
\includegraphics{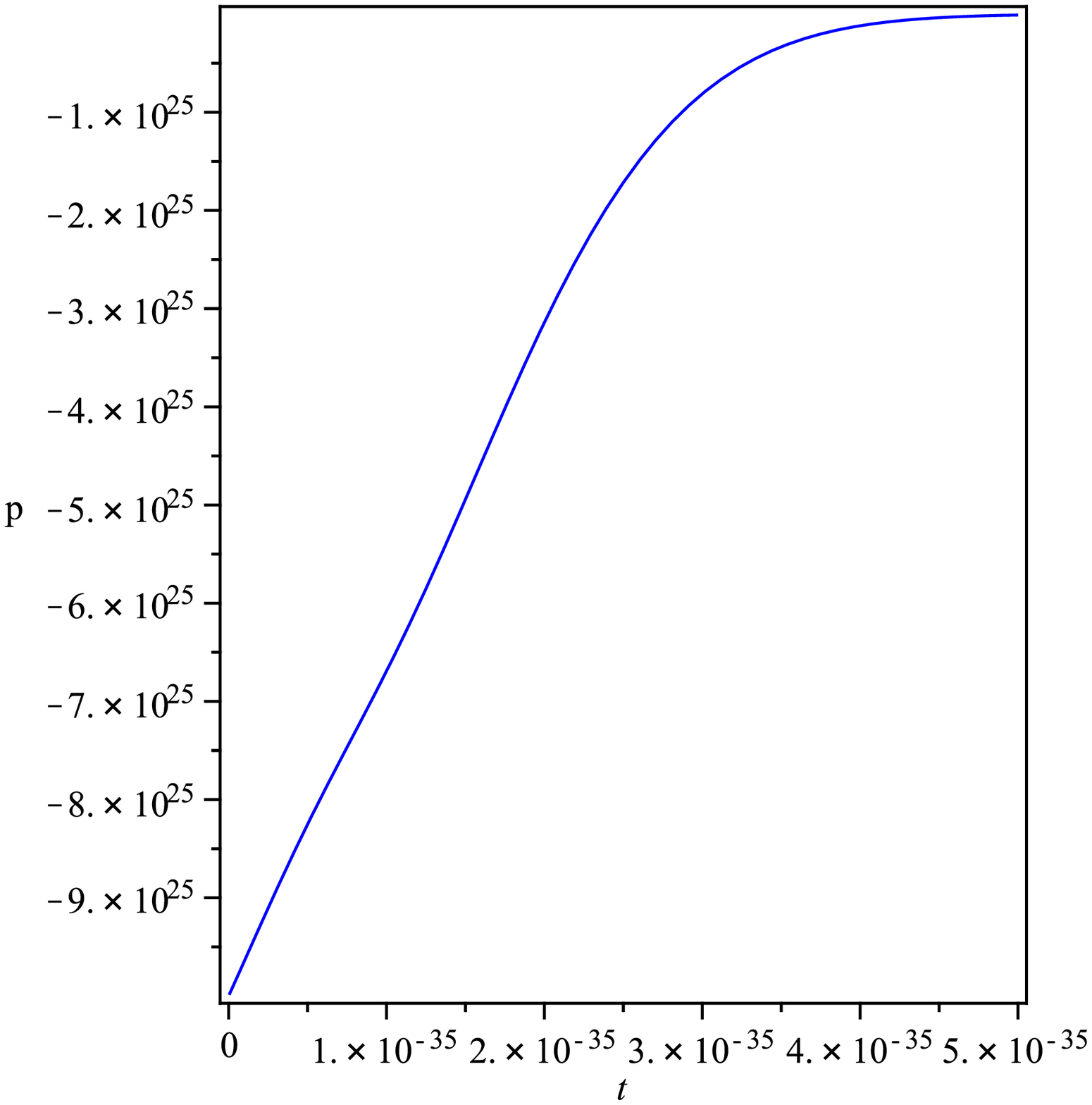} \vspace{5cm}\includegraphics{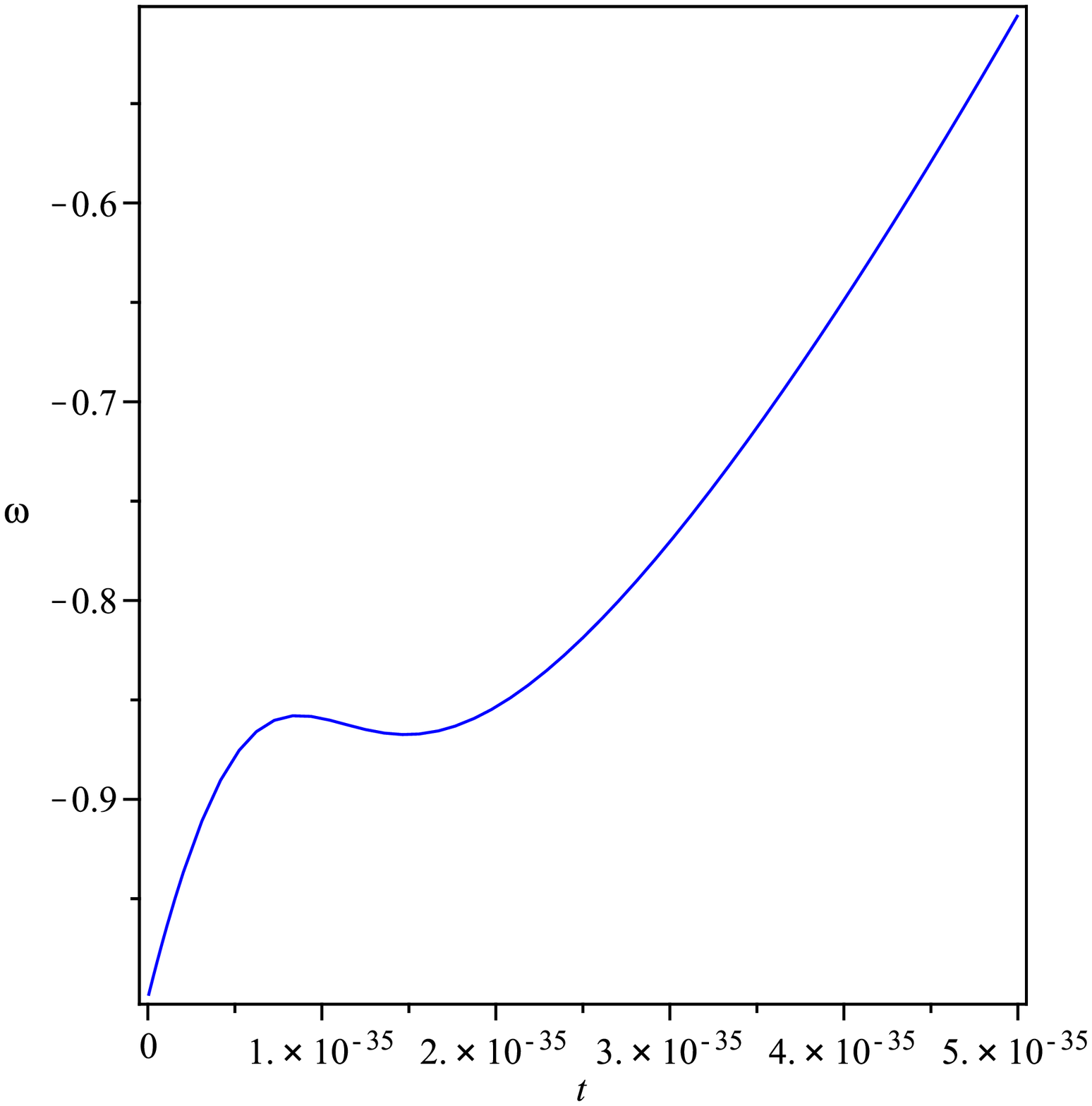}\vspace{2 cm}
\end{center}
\vspace{0cm} \caption{\small {a) Evolution of  pressure in a
noncommutative spacetime with entropic gravity. The pressure is
negative which provides the condition for a successful inflation. b)
Evolution of the equation of state parameter in a noncommutative
spacetime with entropic gravity. Evidently, this model gives an
accelerating behavior ( $\omega<-\frac{1}{3}$) in early stage of the
universe evolution.}}
\end{figure}

\section{Perturbations}
The perturbed flat FRW metric in the longitudinal gauge has the form
\begin{equation}
ds^2=-(1+2\Phi)dt^2+a^2(1-2\Psi)\gamma_{lm}dx^{l}dx^{m}\,
\end{equation}
where $\gamma_{lm}$ is the Euclidian metric and $l\,,\, m=1,2,3$.
Generally, the perturbations are described by $\Phi$ (the Bardeen
potential) and $\Psi$, which both are functions of space and time.
Nevertheless, in the absence of noncommutative stress one could set
$\Phi=\Psi$. The perturbed Einstein equations are
 \begin{equation}
3H(H\Phi+\dot{\Psi})-\frac{1}{a^{2}}\nabla^{2}\Psi=-4\pi G\,
\delta\rho
\end{equation}
\begin{equation}
H\Phi+\dot{\Psi}=-4\pi G\, \delta q
\end{equation}
\begin{equation}
\ddot{\Psi}+3H\dot{\Psi}+H\dot{\Phi}+(2\dot{H}+3H^{2})\Phi-\frac{1}{3a^{2}}\nabla^{2}(\Psi-\Phi)=4\pi
G c^{2}_{s}\, \delta\rho
\end{equation}
where $\delta T_{0}^{0}=-\delta\rho$, $\delta T_{i}^{0}=\delta
q_{i}$ , $\delta T_{j}^{i}= c^{2}_{s}\delta\rho\delta_{j}^{i}$ and
$c_{s}$ is the speed of sound given by
$c^{2}_{s}=\frac{\dot{p}}{\dot{\rho}}$\,. Vanishing of the
anisotropic stress means that we can set $\Phi=\Psi$ in which case.
Then, equations (31) and (33) could be combined to give a single
equation for the evolution of metric potential $\Phi$ in the
momentum space
\begin{equation}
\ddot{\Phi}_{k}+(4+3c^{2}_{s})H\dot{\Phi_{k}}+\Bigg(\frac{k^2c^{2}_{s}}{a^{2}}+2\dot{H}+3H^{2}(1+c^{2}_{s})\Bigg)\Phi_{k}=0\,.
\end{equation}
Using the metric potential $\Phi$ , the power spectrum at the time
of Hubble crossing , $k=aH$ then is given by
\begin{equation}
P_{\Phi_{k}}=\frac{k^{3}}{2\pi^{2}}\langle\Phi^{2}_{k}\rangle\,.
\end{equation}
The curvature perturbation on slices of uniform energy density,
$\zeta$, can be calculated from $\Phi$ using the relation
\begin{equation}
\zeta\equiv\Phi-\frac{H}{\dot{H}}(\dot{\Phi}+H\Phi)\,.
\end{equation}
Using equation (26), the explicit form of the Hubble parameter in
this model is

$$H=\frac{1}{6}\,\Bigg\{ 65536\,\pi \,\sqrt
{3G\rho_{0}\,\theta}\exp\Big[\frac{1}{24}(-3\,{t}^{2}+32\,\pi
\,\sqrt {3}\sqrt {G\rho_{0}\, \theta}) {\rm erf}({\frac {t}{2\sqrt
{2\theta}}})\Big]\sqrt {2}{\theta}^{9/2}$$
$$+\rm
e^{{\frac {t}{\sqrt {2\theta}}}}\Bigg[240\sqrt { 2\pi}\,\sqrt
{\theta}{t}^{4}+48\,{t}^{5}\sqrt {\pi }+600\,\theta\,{\rm Ei}(
1,{\frac {t}{2 \sqrt {2\theta}}}) {t}^{3}+75\,\sqrt
{2\theta}{t}^{4}{\rm Ei}( 1,{\frac {t}{2\sqrt {2\theta}}})\Bigg]
$$

$$-{{\rm e}^{{\frac {t}{2\sqrt {2
\theta}}}}}\Bigg[108\,\theta\,{t}^{3}+768\,{\theta}^{2}t+336\,{\theta}^{3/2}{t}^{2}\sqrt
{2}+768\,{\theta} ^{5/2}\sqrt {2}\Bigg]\Bigg\}$$

$$\times {\frac
{1}{\sqrt {\theta}}}\Bigg\{ 16384\,{ \theta}^{9/2}\sqrt {\pi
}\exp\Big({\frac{4\sqrt {3}}{3}\pi \,\sqrt {G\rho_{0}\,\theta}
{{\rm erf}({\frac {t}{2\sqrt {2\theta}}})}}\Big)+8{t}^{5}\sqrt
{2}{{\rm e}^{{\frac {t }{\sqrt {2\theta}}}}}\sqrt {\pi }+25\,{\rm
Ei}( 1,{ \frac {t\sqrt {2}}{2\sqrt {2\theta}}} ) {t}^{4}{{\rm
e}^{{ \frac {t}{2\sqrt {\theta}}}}}\sqrt {\theta}$$
\begin{equation}
+{{\rm e}^{{\frac {t}{2\sqrt
{2\theta}}}}}\Bigg[56\,{t}^{2}{\theta}^{3/2}+14\,{t}^{3}\sqrt {2}
\theta+384\,{ \theta}^{5/2}+32\,t\sqrt
{2}{\theta}^{2}\Bigg]\Bigg\} ^{-1}\,.
\end{equation}
Given the extremely complicated form of the Hubble parameter and the
speed of sound, equation (34) cannot be solved analytically.
Instead, we have done a numerical analysis and plotted the evolution
of the metric potential $\Phi$ and curvature perturbation. To doing
so, we have used the values $\theta=10^{-40}cm$ and
$\rho_{0}=10^{26}$. Figures 2 and 3 are results of our numerical
analysis. The crucial test for any inflationary model is their
ability to produce almost scale invariant spectrum of scalar
perturbations. As it has been shown previously [16],
noncommutativity generates a small deviation from scale invariance
in inflationary scenarios. Also it is expected that nongaussianity
should be greater in these models than the usual inflaton field
models [17].  To be a realistic model of the early universe and also
to test whether or not our model is consistent with recent
observational data, we define the slow-roll parameters as usual
$$\epsilon\equiv-\frac{\dot{H}}{H^{2}}$$
\begin{equation}
\eta\equiv\frac{\dot{\epsilon}}{H\epsilon}\,\,.
\end{equation}
We assume also that as usual the scalar spectral index is given by
the following relation
\begin{equation}
n_{s}-1\simeq -6\epsilon+2\eta.
\end{equation}

Evolution of the slow roll parameters $\epsilon$ and $\eta$ versus
the cosmic time are given in figure 2 (left panel). Both parameters
are small at the time of inflation, leading to an almost scale
invariant, slightly red tilted spectrum as depicted in the right
panel of this figure. Also figure 3 gives evolution of the metric
potential $\Phi$ (dashed line) and curvature perturbation $\zeta$ (
solid line) versus the cosmic time. $\zeta$ is almost constant at
the end of inflation while $\Phi$ approaches order of unity at that
time. We note that our adopted value of $\rho_{0}=10^{26}$ is chosen
so that a sufficient number of e-foldings to solve standard model
problems is guaranteed.

\begin{figure}[htp]
\begin{center}
\includegraphics{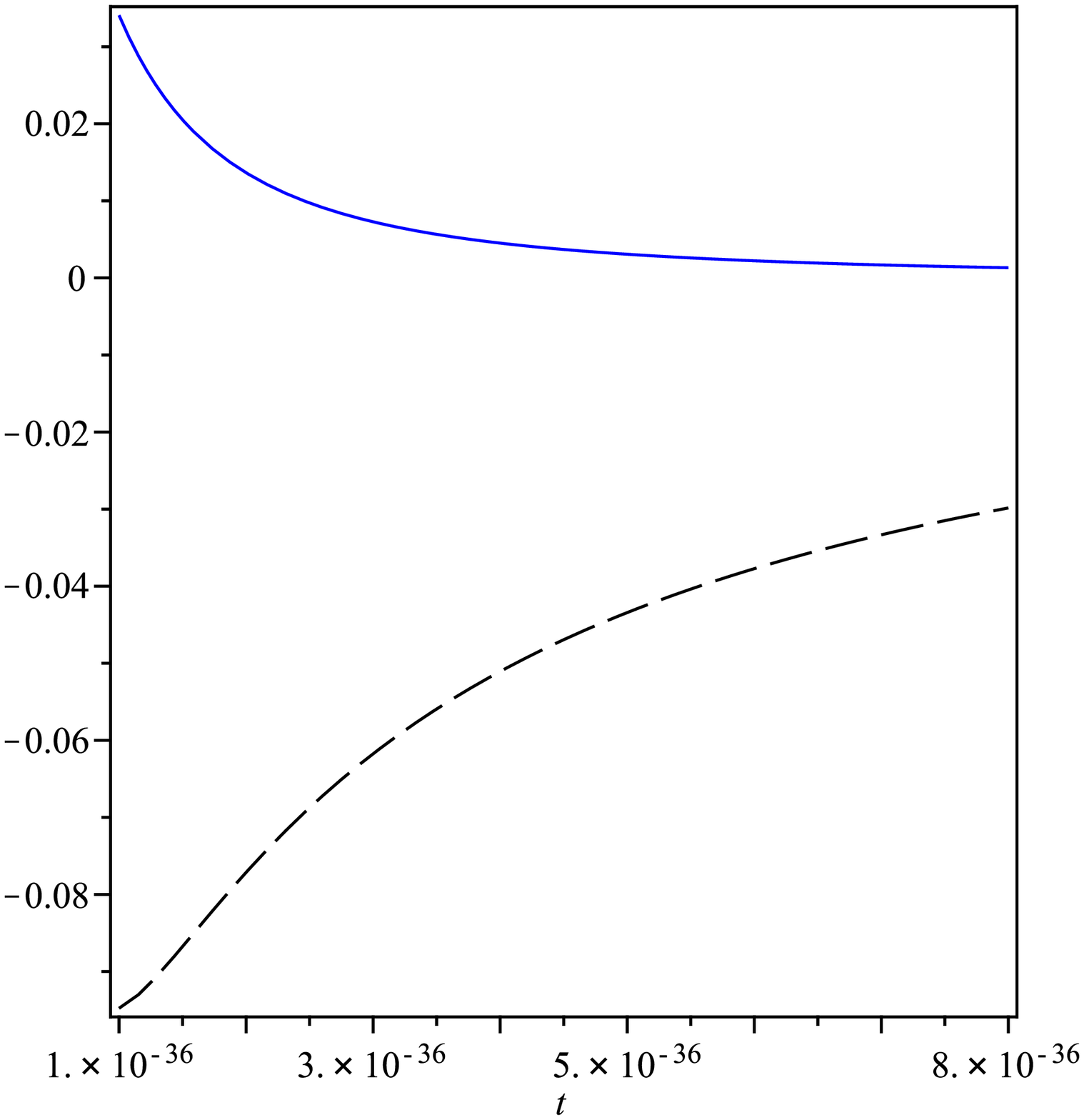} \vspace{5cm}\includegraphics{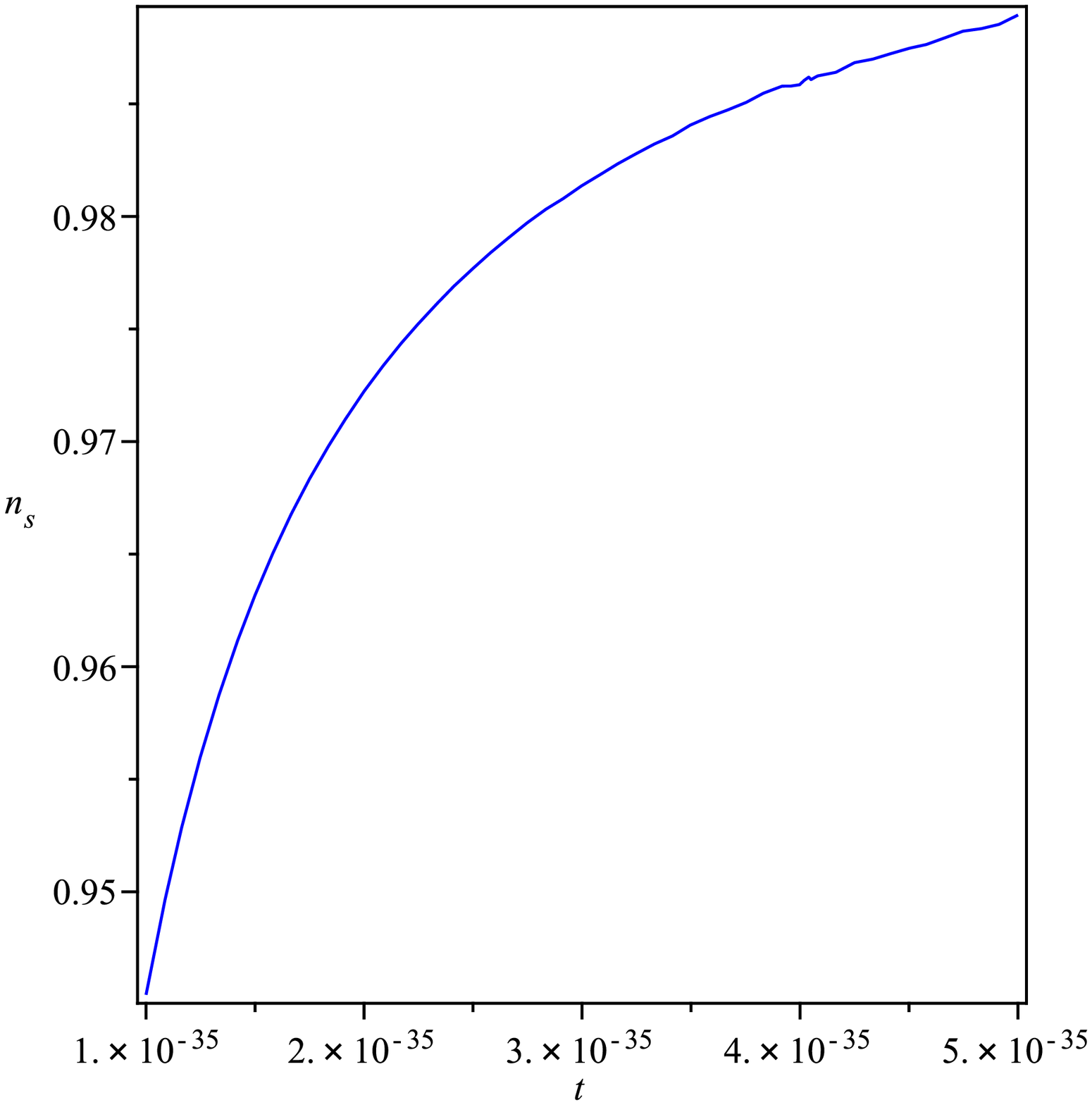}\vspace{2 cm}
\end{center}
\vspace{0cm} \caption{\small {a) Evolution of  the slow roll
parameters $\epsilon$ (solid line) and $\eta$ (dashed line) versus
the cosmic time. Both parameters are small at the time of inflation
leading to an almost scale invariant spectrum. b) Evolution of the
scalar spectral index versus the cosmic time. Due to value and sign
of slow roll parameters, $n_{s}$ derives towards unity at the end of
inflation and is slightly red tilted.}}
\end{figure}
\begin{figure}[htp]
\begin{center}
\includegraphics{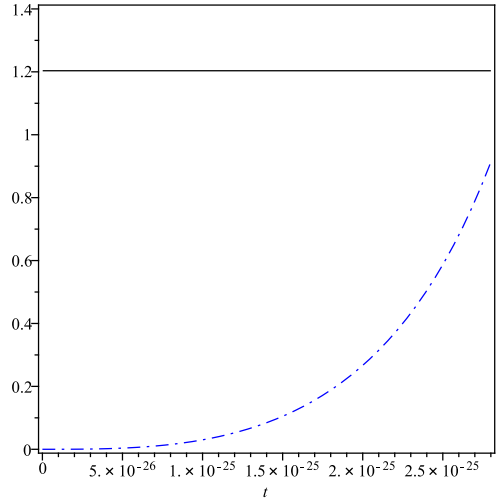}
\end{center}
\vspace{6 cm} \caption{\small {Evolution of the metric potential
$\Phi$ (dashed line) and curvature perturbation $\zeta$ (solid line)
versus the cosmic time. $\zeta$ is almost constant at the end of
inflation while $\Phi$ approaches order of unity at that time.}}
\end{figure}

\section{Conclusion}
The relation between spacetime noncommutativity and thermodynamics
has been known previously and specifically blackhole thermodynamics
in noncommutative spaces has been studied extensively [18].
Noncommutativity encodes microscopic degrees of freedom of the
quantum spacetime and in this sense, a noncommutative manifold will
be equivalent to a thermodynamical system with a temperature given
by noncommutative parameter $\theta$. Considering these facts, it
seems that there is a deep connection between noncommutativity and
thermodynamical description of gravity proposed by Verlinde.
Furthermore, both these theories rely on the existence of a
fundamental unit of length (Planck or string length) and are similar
in this respect too.

Generally it is believed that cosmic microwave background embodies
the effects of trans-Planckian physics [19-25]. Noncommutative
inflation is a process which imprints these effects to CMB, so there
have been various attempts at constructing noncommutative
inflationary models to test the results against observations of CMB.
One approach is to use relation (5) for space-space coordinates
[19,20,26,27] and to construct a noncommutative field theory on the
spacetime manifold by replacing ordinary product of fields by
Weyl-Wigner-Moyal $\ast$-product. Another approach is to incorporate
the fundamental noncommutativity of spacetime into inflationary
models via a generalized uncertainty principle (GUP)[28] (see also
[16]). Coherent state picture of noncommutativity also has been used
to construct inflationary models in both 4D and extra-dimensional
scenarios [15, 29].

In this paper we used the coherent state picture of noncommutativity
to determine the explicit entropy-area relation in order to
determine modified Friedmann equation in a universe governed by
entropic gravity. In this setup, negative pressure of noncommutative
energy density in very early universe derives it toward an
inflationary era. The dynamics of this inflationary epoch and
generation of scale invariant scalar perturbations have been
studied. In this model, the slow-roll parameters are small at the
time of inflation, leading to an almost scale invariant, slightly
red tilted spectrum.\\

{\bf Acknowledgment}\\
This work has been supported partially by Research Institute for
Astronomy and Astrophysics of Maragha, IRAN.


\begin{thebibliography}{12}
\bibitem{1}
E. P. Verlinde, [arXiv:1001.0785].

\bibitem{2}
T. Jacobson, Phys. Rev. Lett. \textbf{75}, 1260 (1995).

\bibitem{3}
W. G. Unruh, Phys. Rev. D \textbf{14}, 870 (1976).

\bibitem{4}
R. G. Cai, L. M. Cao and N. Ohta, Phys. Rev. D \textbf{81}, 061501
(2010), [arXiv:1001.3470].

\bibitem{5}
A. Sheykhi, Phys. Rev. D \textbf{81}, 104011 (2010),
[arXiv:1004.0627]

\bibitem{6}
M. Li and Y. Wang, Phys. Lett. B \textbf{687}, 243 (2010), [arXiv:1001.4466] \\
S. W. Wei, Y. X. Liu and Y. Q. Wang, [arXiv:1001.5238] \\
Y. Ling and J. P. Wu, JCAP \textbf{1008}, 017 (2010), [arXiv:1001.5324] \\
D. A. Easson, P. H. Frampton and G. F. Smoot, [arXiv:1002.4278]\\
U. H. Danielsson, [arXiv:1003.0668] \\
N. Afshordi, [arXiv:1003.4811]\\
A. Sheykhi, Phys. Rev. D \textbf{81}, 104011 (2010), [arXiv:1004.0627]\\
A. Sheykhi and S. H. Hendi, [arXiv:1011.0676]\\
M. Li and Y. Pang, Phys. Rev. D\textbf{ 82}, 027501 (2010), [arXiv:1004.0877]\\
T. Qiu, [arXiv:1004.1693] \\
R. Casadio and A. Gruppuso, [arXiv:1005.0790] \\
H. Wei, [arXiv:1005.1445]\\
W. Gu, M. Li and R. -X. Miao, [arXiv:1011.3419]\\
T. S. Koivisto, D. F. Mota and M. Zumalacarregui,
[arXiv:1011.2226]\\
S. H. Hendi and A. Sheykhi, [arXiv:1012.0381].

\bibitem{7}
D. A. Easson, P. H. Frampton and G. F. Smoot, [arXiv:1003.1528] \\
M. Li and Y. Pang, Phys. Rev. D \textbf{82}, 027501 (2010), [arXiv:1004.0877]\\
Y. F. Cai, J. Liu and H. Li, Phys. Lett. B \textbf{690}, 213 (2010), [arXiv:1003.4526]\\
Y. F. Cai and E. N. Saridakis, [arXiv:1011.1245].

\bibitem{8}
M. R. Douglas and N. A. Nekrasov, Rev. Mod. Phys. {\bf 73}
977, (2001)\\
R. J. Szabo, Phys. Rept. {\bf 378}, 207 (2003) \\
N. Seiberg and E. Witten, JHEP {\bf9909}, 032 (1999) \\
A. Connes and M. Marcolli,  [arXiv:math.QA/0601054] \\
A. Connes, J. Math. Phys. {\bf 41}, 3832 (2000)\\
A. Konechny and A. Schwarz, Phys. Rept. {\bf 360}, 353 (2002)\\
M. Chaichian {\it et al}, Eur. Phys. J. C {\bf29}, 413 (2003)\\
A. Micu and  M. M. Sheikh-Jabbari,  JHEP {\bf 0101}, 025
(2001)\\
M. Panero, JHEP \textbf{0705}, 082 (2007).

\bibitem{9}
G. Veneziano, Europhys. Lett. {\bf2}, 199 (1986)\\
D. Amati, M. Ciafaloni and G. Veneziano, Phys. Lett. B
{\bf197}, 81 (1987) \\
D. Amati, M. Ciafaloni and G. Veneziano, Int. J. Mod. Phys. A
{\bf3}, 1615 (1988) \\
D. Amati, M. Ciafaloni and G. Veneziano,  Nucl. Phys. B
{\bf347},530 (1990) \\
D. J. Gross and P. F. Mende,  Nucl. Phys. B {\bf 303}, 407
(1988)\\
D. Amati, M. Ciafaloni and G. Veneziano, Phys. Lett. B {\bf 216}, 41
(1989).

\bibitem{10}
P. Nicolini, Phys. Rev. D \textbf{82}, 044030 (2010),
[arXiv:1005.2996].
\bibitem{11}
A. Strominger and C. Vafa, Phys. Lett. B \textbf{379}, 99 (1996);\\
E. Halyo, B. Kol, A. Rajaraman and L. Susskind, Phys. Lett. B
\textbf{401},
15 (1997);\\
S. N. Solodukhin, Phys. Rev. D \textbf{57}, 2410 (1998)\\
L Modesto, Class. Quant. Grav. \textbf{23}, 5587-5602 (2006) [arXiv:0509078]; \\
L Modesto, [arXiv:0811.2196 [gr-qc]];\\
J. Zhang,Phys.Lett. B 668 (2008) 353;\\
R. Banerjee and B. R. Majhi, Phys. Lett. B \textbf{662}, 62 (2008);\\
R. Banerjee and B. R. Majhi, JHEP \textbf{0806} (2008) 095;
\bibitem{12}
A. Smailagic and E. Spallucci, J. Phys. A \textbf{37}, 1
(2004), [Erratum-ibid: J. Phys. A {\bf37}, 7169 (2004)]\\
A. Smailagic and E. Spallucci, J. Phys. A \textbf{36}, L467 (2003) \\
A. Smailagic and E. Spallucci, J. Phys. A \textbf{36}, L517 (2003).
\bibitem{13}
R. Cai, L. Cao, N. Ohta,  Phys. Rev. D {\bf81}, 084012 (2010) [arXiv:1002.1136];\\
F. Shu, Y. Gong,[arXiv:1001.3237]
\bibitem{14}
T. Padmanabhan, Class. Quant. Grav. {\bf21}, 4485 (2004)
[arXiv:gr-qc/0308070]
\bibitem{15}
M. Rinaldi, [arXiv:0908.1949].

\bibitem{16}
K. Nozari and S. Akhshabi, Int. J. Mod. Phys. D \textbf{19}, 513
(2010), [arXiv:0910.2808]\\
Q. G. Huang, M. Li, Nucl. Phys. B \textbf{713}, 219 (2005)\\
Q. G. Huang, M. Li, Nucl. Phys. B \textbf{755}, 286 (2006) \\
S. Koh, R.H. Brandenberger, JCAP \textbf{0706}, 021 (2007).

\bibitem{17}
K. Fang, B. Chen and  W. Xue, Phys. Rev. D\textbf{77}, 063523
(2008).

\bibitem{18}
P. Nicolini, A. Smailagic and E. Spallucci, ESA Spec. Publ.
\textbf{637}, 11.1 (2006), [arXiv:hep-th/0507226]\\
P. Nicolini, J. Phys. A \textbf{38}, L631 (2005), [arXiv:hep-th/0507266]\\
P. Nicolini, A. Smailagic and E. Spallucci, Phys. Lett. B
\textbf{632}, 547 (2006), [arXiv:gr-qc/0510112]\\
S. Ansoldi, P. Nicolini, A. Smailagic and E. Spallucci, Phys. Lett.
B \textbf{645}, 261 (2007), [arXiv:gr-qc/0612035]\\
E. Spallucci, A. Smailagic and P. Nicolini, Phys. Lett. B \textbf{670}, 449 (2009), [arXiv:0801.3519]\\
Y. S. Myung and M. Yoon, Eur. Phys. J. C \textbf{62}, 405 (2009), [arXiv:0810.0078]\\
M. I. Park, Phys. Rev. D \textbf{80}, 084026 (2009),
[arXiv:0811.2685]\\
R. Garattini and F. S. N. Lobo, Phys. Lett. B \textbf{671}, 146
(2009), [arXiv:0811.0919]\\
P. Nicolini and E. Spallucci, Class. Quant. Grav. \textbf{27}, 015010 (2010), [arXiv:0902.4654] \\
I. Arraut, D. Batic and M. Nowakowski, Class. Quant. Grav.
\textbf{26}, 245006 (2009) [arXiv:0902.3481]\\
I. Arraut, D. Batic and M. Nowakowski, J. Math. Phys. \textbf{51},
022503 (2010), [arXiv:1001.2226]\\
D. Batic and P. Nicolini, [arXiv:1001.1158]\\
W. H. Huang, [arXiv:1003.1040]\\
A. Smailagic and E. Spallucci, Phys. Lett. B \textbf{688}, 82
(2010), [arXiv:1003.3918 [hep-th]].

\bibitem{19}
R. Easther, B. R. Greene, W. H. Kinney and G. Shiu, Phys. Rev. D
\textbf{64}, 103502 (2001).

\bibitem{20}
R. Easther, B. R. Greene, W. H. Kinney and G. Shiu, Phys. Rev. D
\textbf{67}, 063508 (2003).

\bibitem{21}
R. Easther, B. R. Greene, W. H. Kinney and G. Shiu, Phys. Rev. D
\textbf{66}, 023518 (2002).

\bibitem{22}
N. Kaloper, M. Kleban, A. E. Lawrence and S. Shenker, Phys. Rev. D
\textbf{66}, 123510 (2002).

\bibitem{23}
L. Bergstrom, U. H Danielsson, JHEP \textbf{07},038 (2002).

\bibitem{24}
J. Martin and R. Brandenberger, Phys. Rev. D \textbf{68}, 0305161
(2003).
\bibitem{25}
O. Elgaroy, S. Hannestad,  Phys. Rev. D \textbf{62}, 041301 (2000)
041301.

\bibitem{26}
S. Chu, B. R. Greene and G. Shiu, Mod. Phys. Lett. A \textbf{16},
2231 (2001)\\
F. Lizzi, G. Mangano, G. Miele and M. Peloso, JHEP \textbf{0206},
049
(2002) [arXiv:hep- th/0203099]\\
S. F. Hassan and M. S. Sloth, Nucl. Phys. B \textbf{674}, 434
(2003), [arXiv:hep-th/0204110].
\bibitem{27}
R. Brandenberger and P. M. Ho, Phys. Rev. D \textbf{66}, 023517
(2002).
\bibitem{28}
S. Alexander and J. Magueijo, [arXiv:hep-th/0104093]\\
S. Alexander, R. Brandenberger and J. Magueijo, Phys. Rev. D
\textbf{67}, 081301 (2003) [arXiv:hep-th/0108190]\\
S. Koh, Mod. Phys. Lett. A \textbf{23}, 1598 (2008).
\bibitem{29}
K. Nozari and S. Akhshabi, Phys. Lett. B {\bf683}, 186 (2010), [arXiv:0911.4418]\\
K. Nozari and S. Akhshabi, [arXiv:1004.5007].
\end{thebibliography}
\end{document}